\documentstyle{article}

\newtheorem{theorem}{Theorem}
\newtheorem{definition}{Definition}
\newtheorem{lemma}{Lemma}

\newcommand{\putaddress}{\setcounter{footnote}{1}
    \footnotetext{Department of Computer Science,
University of Texas at El Paso,
El Paso, TX 79968, USA
}
    \setcounter{footnote}{2}
    \footnotetext{Department of Mathematics, SPb UEF,
    Griboyedova 30/32, 191023, St-Petersburg,
    Russia (address for correspondence)}
    \setcounter{footnote}{3}
    \footnotetext{Division of Mathematics,
    Istituto per la Ricerca di Base,
    I-86075, Monteroduni (IS), Molise, Italy}}

\newcommand{\riem}{R}
\newcommand{\scrv}{R}
\newcommand{\rnum}{{\bf R}}

\title{An operationalistic reformulation of Einstein's equivalence
principle}

\author{Vladik Kreinovich\footnotemark[1],
R.R.Zapatrin\footnotemark[2]{\makebox[0.4em]{}}
\footnotemark[3]
}

\date{}

\begin{document}
\unitlength=1mm

\maketitle
\putaddress

\begin{abstract}
The Einstein's equivalence principle is formulated in terms of
the accuracy of measurements and its dependence of the size of
the area of measurement. It is shown that different refinements
of the statement 'the spacetime is locally flat' lead to different
conculsions about the spacetime geometry.
\end{abstract}

\section{Introduction}

Analyzing gravitational phenomena Einstein used the following
postulate (which he called equivalence principle): what ever
measurements we perform inside some spacetime region we cannot
distinguish between the case when there is a homogeneous
gravitational field and the case when all bodies in this region
have constant acceleration with respect to some inertial frame.
(And since any field can be considered homogeneous in a small
enough region, this principle can be applied to a neighborhood of
any point).

Einstein concluded from this principle that the spacetime
metric is pseudo-Riemannian and in absence of all other fields but
gravity the test particles are traveling along geodesics of this
metric \cite{einstein}.

Yet V.A.Fock \cite{fock} noticed that this formulation is not
exact enough: according to general relativity , the presence of
gravitation means spacetime to be curved, i.e. curvature tensor is
nonzero, $\riem_{ijkl} \neq 0$.  This is valid in any frame, in
particular in a uniformly accelerated one.  Hence in presence of
gravity $\riem_{ijkl} \neq 0$ while in uniformly accelerated frame
$\riem_{ijkl} = 0$, and this can be distinguished experimentally by
emitting a "cloud" of a particles endowed with clocks to various
directions with various speeds.  With a help of the clocks one can
determine the proper time $ds$ along every trajectory and then
calculate the metric.  By numerical differentiation of the metric we
can obtain the values of $\riem_{ijkl}$ and then compare all them
with zero.

That is why most of authors do postulate the Riemannian metric
within the strict mathematical account of general relativity.  We
would like to give here more profound and at the same time more
strict mathematical grounds for this fact.

The main drawback of traditional definition of Riemannian
geometry of spacetime is that it is formulated in terms of length
({\em viz.}\/ proper time) of idealized infinitely small intervals
rather than real ones of finite size.  Besides that, this definition
demands length of space-like intervals to be determined which is not
desirable from the operationalistic point of view.  Some authors
(see, {\em e.g.} \cite{c4}) give the equivalent definition
including only proper time along finite parts of time-like curves.
The postulate the metric to be Riemannian in the sense of this
definition. This is more operationalistic but yet not motivated
physically.

In our paper we show that one can reformulate the initial Einstein's
equivalence principle in such a way that both Riemannian metric of
spacetime and the geodesic motion of test particles will be obtained
from it.

Considering only the gravitational field this result is of few
interest, but the question becomes essential in presence of non
gravitational fields - it stipulates the choice of covariant analogue
of an equation.  For example in \cite{c7} one asserts that
conformally invariant scalar field equations $\Box\phi + 1/6\,\scrv
\phi=0$ come into contradiction with equivalence principle since
they contain scalar curvature $\scrv$ (a more detailed analysis of
this issue can be found in \cite{c8}). Nevertheless, such
reasonings does not seem convincing: for example, usual Maxwell
equations in curved space contain explicitly the curvature, but
they undoubted by agree with equivalence principle.  However if we
twice differentiate both parts of the equation and exchange
$F_{kl;ij}$ by $F_{kl;ji}+{{\riem_{ij}}^p}_l F_{kp} +
{{\riem_{ij}}^p}_k F_{pl}$ we obtain equations containing the
curvature explicitly. Thus the presence of curvature tensor in an
equation does not mean at all the violation of equivalence
principle. The formulation of equivalence principle proposed below
allows to solve this problem in a physically meaningful and
mathematically strict way.

The idea of our reformulation is the following.  The Fock's
experiment with the cloud of particles described above is idealized
since all real measurements have their errors. Therefore all the
values calculated via these measurements, in particular, the
curvature, have their errors too.  Thus if the error of so calculated
curvature tensor will be great enough (of the order of the curvature
itself) it would not be possible to determine whether the genuine
value of curvature tensor is equal to zero or not. In the meantime,
the Einstein's principle claims that any {\em real} (rather than
exact) measurement performed in small enough region will not allow us
to distinguish real (possibly curved) space from flat one.

\section{Mathematical formulation}\label{s2}

Begin with a formalization of basic notions.

\begin{definition}\label{d1} A spacetime region $M$ is called
$\epsilon${\sc -small} with respect to some fixed frame iff for any
$i$ and any $a,b\in M$ $$ \vert x^i(a) - x^i(b)\vert < \epsilon $$
\end{definition}

This definition depends on coordinate frame; however the
reformulation of equivalence principle based on this definition turns
out not to depend on frame.

The spacetime properties determine the
relative movement of uncharged particles.  There are devices to
measure coordinates and other kinematic features of the motion: time,
velocity (e.g. using the Doppler effect), acceleration etc.  However
one mostly measure time or length (e.g. Doppler measurement of
velocity contains determining frequency - the time interval between
neighbor maxima). So, further we shall consider only time and
length measurement. Clearly the measuring of small intervals of time
and length can be performed with smaller absolute error. Let us
denote by $\lambda(\epsilon)$ the error of our measurements in the
$\epsilon$-small region.

\begin{definition}\label{d2} A {\sc spacetime} is a triple
$(M,\Gamma,\tau)$ with $M$ -- a smooth manifold, $\Gamma$ -- a
family of smooth curves on $M$ (trajectories of test particles) and
for any $\gamma \in \Gamma$ a smooth function $\tau:\gamma \times
\gamma \rightarrow \rnum$ (the proper time along $\gamma$) is
determined such that

\[
\tau(a,c) = \tau(a,b) + \tau(b,c)
\]

\noindent whenever

\[
\gamma^{-1}(a) < \gamma^{-1}(b) < \gamma^{-1}(c)
\]

\end{definition}

The {\sc flat} spacetime is a triple $(M_0,\Gamma_0,\tau_0)$ where
$M_0 = \rnum^4$, $\Gamma_0$ is set of all time-like straight lines,
$\tau_0$ is Minkowski metric.

\begin{definition}\label{d3}  A spacetime $(M,\Gamma,\tau)$
is called $\lambda$-flat if for any point $m\in M$ there exists
such a frame that for a sufficiently small $\epsilon > 0$ all
coordinate and time measurements in any $\epsilon$-small region of
$M$ coincide (up to an error $\le \lambda(\epsilon)$) with the
analogous result in the flat spacetime.
\end{definition}

The final formulation of the equivalence principle must not of course
depend on accessible devices ({\em i.e.}\/ the kind of the function
$\lambda$). Thus, instead of a single function we must operate with
a class of such functions $\Lambda = \{\lambda\}$. We shall assume
possible refinement of any measurement, namely $\Lambda$ together
with every $\lambda$ is assumed to contain also
the function $k\lambda$ for every $0 < k < 1$.

\begin{definition}\label{d4}  A spacetime is siad to be
$\Lambda$-flat if it is $\lambda$-flat for all $\lambda \in
\Lambda$.
\end{definition}

The formulation of the equivalence principle we propose is the
following:

\[
\hbox{\em the spacetime is} \; \lambda\hbox{\em -flat}
\]

If we exclude degenerate cases ($\Lambda$ is too great and Fock's
reasoning is valid or $\Lambda$ is too small so that
$\Lambda$-flatness implies nothing) the proposed formulation yields
us a basis for Riemannian metrics and geodesic motion.

\section{Main results}\label{sres}

\begin{theorem}\label{th1} For any class of functions
$\Lambda:\rnum^+ \rightarrow \rnum^+$ such that $\lambda \in
\Lambda$ implies $k \lambda \in \Lambda$ for any positive $k\le 1$
one of the following statements is valid:  \end{theorem}

\begin{itemize}
\item[A.] Any spacetime is $\Lambda$-flat.
\item[B.] Only pseudo-Riemannian spacetime are $\Lambda$-flat and
the class of curves $\Gamma$ is arbitrary.
\item[C.] Only pseudo-Riemannian spacetimes are $\Lambda$-flat
and $\Gamma$ is the set of timelike geodesics.
\item[D.] Only flat spacetime is $\Lambda$-flat.
\end{itemize}

The proof will be organized according to the following plan:

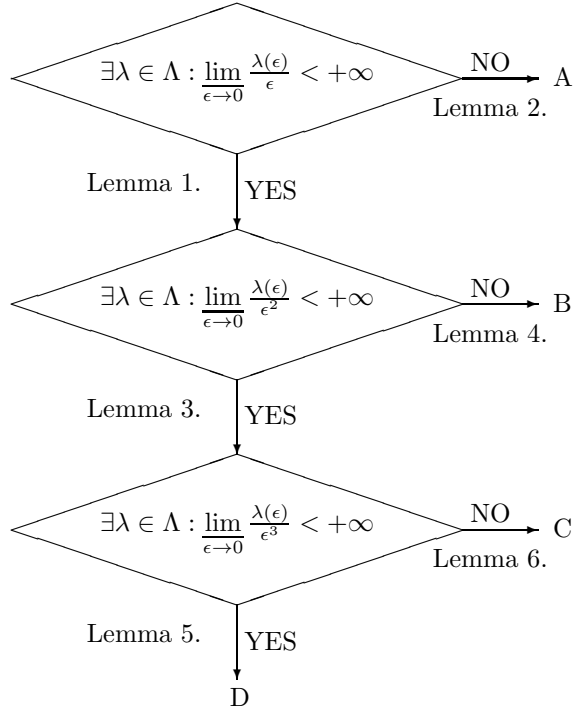
\begin{picture}(120,100)(0,-10)
\multiput(10,15)(0,30){3}{\line(3,-1){30}}
\multiput(10,15)(0,30){3}{\line(3,1){30}}
\multiput(70,15)(0,30){3}{\line(-3,-1){30}}
\multiput(70,15)(0,30){3}{\line(-3,1){30}}
\multiput(40,5)(0,30){3}{\vector(0,-1){10}}
\put(20,0){\mbox{Lemma \ref{l5}.}}
\put(20,30){\mbox{Lemma \ref{l3}.}}
\put(20,60){\mbox{Lemma \ref{l1}.}}
\multiput(41,-1)(0,30){3}{\mbox{YES}}
\multiput(70,15)(0,30){3}{\vector(1,0){10}}
\put(66,10){\mbox{Lemma \ref{l6}.}}
\put(66,40){\mbox{Lemma \ref{l4}.}}
\put(66,70){\mbox{Lemma \ref{l2}.}}
\multiput(71,16)(0,30){3}{\mbox{NO}}

\put(40,15){\makebox(0,0){$\exists \lambda \in \Lambda :
\lim\limits_{\overline{\epsilon \to
0}}\frac{\lambda(\epsilon)}{\epsilon^3} < +\infty$}}
\put(39,-8.5){\mbox{D}}
\put(82,14){\mbox{C}}

\put(40,45){\makebox(0,0){$\exists \lambda \in \Lambda :
\lim\limits_{\overline{\epsilon \to
0}}\frac{\lambda(\epsilon)}{\epsilon^2} < +\infty$}}
\put(82,44){\mbox{B}}

\put(40,75){\makebox(0,0){$\exists \lambda \in \Lambda :
\lim\limits_{\overline{\epsilon \to
0}}\frac{\lambda(\epsilon)}{\epsilon} < +\infty$}}
\put(82,74){\mbox{A}}

\end{picture}

\begin{lemma}\label{l1} If there exists $\lambda \in \Lambda$ for
which

\begin{equation}\label{e1}
\lim\limits_{\overline{\epsilon \to
0}}\frac{\lambda(\epsilon)}{\epsilon}\,=\,K\, < +\infty
\end{equation}

\noindent then any $\Lambda$-flat spacetime is pseudo-Riemannian.
\end{lemma}

\paragraph{Proof.} Consider a curve $\gamma \in \Gamma$ and a point
$a\in m$ in a coordinate frame $\{x^i\}$. Let $a$ have the
coordinates $\{x^i_0\}$ in this frame.

In accordance with the definition of $\underline{\lim}$ there exists
such a sequence $\epsilon_n$ that $\lambda(\epsilon_n)/\epsilon_n$
tends to $K$. Let all $\epsilon_n$ be small enough (it can be
assumed with no loss of generality) then for any $n$ all measurements
in the region
$$ \vert x^i - x^i_0 \vert \,<\, \frac{\epsilon_n}{2} $$
\noindent with the error not greater than $\lambda(\epsilon_n)$
coincide with same measurements in flat spacetime, in particular

\[
\vert \delta\tau -\delta\tau_0\vert \,\le\, \lambda(\epsilon_n)
\]

\noindent where

\[
\begin{array}{rcl}
\delta\tau &=& \tau(x^i_0+\delta x^i, x^i_0) \cr
\delta\tau_0 &=& \tau_0(x^i_0+\delta x^i, x^i_0)
\end{array}
\]

\noindent and $\tau,\tau_0$ are metrics along two geodesics both
passing through the region described above.

If $x_0+\delta x$ lies on the frontier of the region then $\vert
\delta x^i \ge C\epsilon_n$ for some  $C = \hbox{const}$ thus

\[
\left\vert \frac{\delta\tau}{\delta x^i} -
\frac{\delta\tau_0}{\delta x^i} \right\vert \,\le\,
\frac{\lambda(\epsilon_n)}{C\epsilon_n}
\]

\noindent therefore

\[
\frac{\delta\tau_0}{\delta x^i} -
\frac{\lambda(\epsilon_n)}{C\epsilon_n} \,\le\,
\frac{\delta\tau}{\delta x^i} \,\le\,
\frac{\delta\tau}{\delta x^i} +
\frac{\lambda(\epsilon_n)}{C\epsilon_n}
\]

So when  $n\to \infty$ (and $\epsilon_n\to 0$)

\[
\frac{\delta\tau_0}{\delta x^i} - \frac{K}{C} \,\le\,
\underline{\lim}\frac{\delta\tau}{\delta x^i} \,\le\,
\overline{\lim}\frac{\delta\tau}{\delta x^i} \,\le\,
\frac{d\tau_0}{dx^i} + \frac{K}{C}
\]

\noindent All the above reasonings are valid for any $k\lambda$ with
$k<1$ hence

\[
\frac{d\tau_0}{dx^i} + \frac{kK}{C} \,\le\,
\underline{\lim}\frac{\delta\tau}{\delta x^i} \,\le\,
\overline{\lim}\frac{\delta\tau}{\delta x^i} \,\le\,
\frac{d\tau_0}{dx^i} + \frac{kK}{C}
\]

Since $k$ can be taken arbitrary small, we have

\[
\lim\frac{\delta\tau}{\delta x^i} \,=\,
\frac{d\tau_0}{dx^i}
\]

\noindent {\em i.e.}\/ in any point in some coordinate frame the
metric of our spacetime coincides with Minkowskian one, that is why
it is pseudo-Riemannian.
\hspace*{\fill}$\Box$\medskip

\begin{lemma}\label{l2} If for any $\lambda \in \Lambda$

\begin{equation}\label{e2}
\underline{\lim}\frac{\lambda(\epsilon)}{\epsilon} \,= \,
+\infty
\end{equation}

\noindent then any spacetime is  $\Lambda$-flat.
\end{lemma}

\paragraph{Proof.} (\ref{e2}) implies $\lim\limits{\epsilon\to 0}
\lambda(\epsilon)/\epsilon = +\infty$. Hence for arbitrary $N$ we
have $\lambda(\epsilon)>N\epsilon$ beginning from some $\epsilon$.
In particular, it is valid for $N>\sup\vert d\tau/dx^i\vert$, hence
$$ \delta\tau < N\delta x^i \le N\epsilon < \lambda(\epsilon) $$
\noindent thus $\vert \delta\tau_0 - \delta\tau\vert <
\lambda(\epsilon)$ for any $\lambda \in \Lambda$.
\hspace*{\fill}$\Box$\medskip

\begin{lemma}\label{l3} If there exists $\lambda \in \Lambda$ such that

\begin{equation}\label{e3}
\underline{\lim}\frac{\lambda(\epsilon)}{\epsilon^2} \,< \,
+\infty
\end{equation}

\noindent then in any $\Lambda$-flat spacetime the set $\Gamma$ is a
set of geodesics.
\end{lemma}

\paragraph{Proof} is similar to that of Lemma \ref{l1}, but uses
the second derivatives of $\tau$:

\[
\left\vert \frac{d^2x^i}{d\tau^2} -
\frac{d^2x^i}{d\tau^2_0} \right\vert \le
\frac{\lambda(\epsilon)}{\epsilon^2}
\]

\noindent hence is some coordinate frame $d^2x^i/d\tau^2 = 0$ thus
$D^2x^i/ds^2 = 0$.
\hspace*{\fill}$\Box$\medskip

\begin{lemma}\label{l4} If for any $\lambda\in\Lambda$

\begin{equation}\label{e4}
\underline{\lim}\frac{\lambda(\epsilon)}{\epsilon^2} \,= \,
+\infty
\end{equation}

\noindent then any pseudo-Riemannian space with any set of
trajectories $\Gamma$ is $\Lambda$-flat.

\end{lemma}

\paragraph{Proof.} Is similar to that of Lemma \ref{l2}. We obtain
that

\[
\lambda(\epsilon) >
\vert D^2x^i/ds^2 \vert =
\vert D^2x^i/ds^2 - 0 \vert =
\vert D^2x^i/ds^2 -D^2x^i_0/ds^2   \vert
\]

\hspace*{\fill}$\Box$\medskip

\begin{lemma}\label{l5} If for some $\lambda\in\Lambda$

\begin{equation}\label{e5}
\underline{\lim}\frac{\lambda(\epsilon)}{\epsilon^3} \,< \,
+\infty
\end{equation}

\noindent then any $\Lambda$-flat spacetime is flat.

\end{lemma}

\paragraph{Proof.} In this case the Fock's reasoning is valid: in
terms of $d^3\tau/dx^3_i$ we can define the curvature tensor with
arbitrary small error.
\hspace*{\fill}$\Box$\medskip

\begin{lemma}\label{l6} If for any $\lambda\in\Lambda$

\begin{equation}\label{e6}
\underline{\lim}\frac{\lambda(\epsilon)}{\epsilon^3} \,= \,
+\infty
\end{equation}

\noindent and for some $\lambda\in\Lambda$

\begin{equation}\label{e7}
\underline{\lim}\frac{\lambda(\epsilon)}{\epsilon^2} \,< \,
+\infty
\end{equation}

\noindent then any pseudo-Riemannian space with the set of
geodesics $\Gamma$ is $\Lambda$- flat.

\end{lemma}

\paragraph{Proof} is carried out likewise. This competes the proof
of the main theorem.
\hspace*{\fill}$\Box$\medskip

The physical meaning of the results obtained is the following:
$\lim\frac{\lambda(\epsilon)}{\epsilon} \,< \,
+\infty$ means the possibility of arbitrary exact measuring
of velocities, $\lim\frac{\lambda(\epsilon)}{\epsilon^2} \,< \,
+\infty$ means the possibility of arbitrary exact measuring
of accelerations, and $\lim\frac{\lambda(\epsilon)}{\epsilon^3} \,<
\, +\infty$ means the possibility of arbitrary exact measuring of
derivatives of accelerations.  So the physical meaning of the
result we obtained is the following: the equivalence principle is
valid only for measuring velocities and accelerations (in any point
they can be turned to zero by corresponding choice of the
coordinate frame) but not valid for derivatives of accelerations
(which correspond to invariant tidal forces).

\section{Some remarks on other fields}\label{srem}

We require the results of all measurements (including trajectories
of particles traveling under other fields) performed in
$\epsilon$-small region (see Def.1) to coincide up to
$\lambda(\epsilon)$ with the results of analogous experiments in
flat spacetime.  We also require it to be so for all functions
$\lambda$ of a class $\Lambda$ containing a function
$\underline{\lim}{\lambda(\epsilon)}/{\epsilon^2} \,< \, +\infty$
satisfying and containing no function $\lambda$ with
$\underline{\lim}{\lambda(\epsilon)}/{\epsilon^3} \,< \, +\infty$

Let us demonstrate that the equivalence principle in this
formulation holds for equation of radiating charged particle and
theories with conformally invariant scalar field and does not hold
for scalar-tensor Brans-Dicke theory.

The equation describing the motion of a {\em radiating charged
particle} contains second derivatives of its velocity $\ddot{u}_i =
d^2u_i/ds^2$ viz. third derivatives of coordinates measured up to
$\lambda(\epsilon)/\epsilon^3$. However
$\lambda(\epsilon)/\epsilon^3$ tends to infinity for any
$\lambda\in\Lambda$, hence the smaller is the region, the greater
is vagueness in determining $\ddot{u}_i$, thus any equation
containing $\ddot{u}_i$ does not contradict our equivalence
principle (including equations containing the curvature
explicitly).

On {\em conformally-invariant scalar field.\/} As we already
mentioned, the only way to measure the curvature is by exploring
its influence on trajectories of particles in accordance with the
formula $\ddot{x}^i = \varphi_{,i}$. Since $\ddot{x}^i$ is measured
up to $\lambda(\epsilon)/\epsilon^3$, the measurement of
$\varphi_{,i}$ has the same value, hence $\Box\varphi =
(\varphi_{,i}^{,i}$ is measured up to
$\lambda(\epsilon)/\epsilon^3$ and in accordance with the preceding
reasoning any equation containing $\Box\varphi$ do not contradict
the operational equivalence principle.

In {\em Brans-Dicke theory} the principle does not hold since
micro-black holes do not move along geodesics there.

\medskip

\section{Locally almost isotropic space is almost uniform: on the
physical meaning of Schur's theorem}\label{s5}

The Schur's theorem asserts that if a space is locally isotropic
(i.e. in any point the curvature tensor has no directions chosen by
some properties)

\begin{equation}\label{e8}
R_{ijkl} \,=\, K(x)(g_{ik}g_{jl} - g_{il}g_{jk})
\end{equation}

\noindent then the space is homogeneous.

\medskip

Since all the real measurements are not exact their results can
provide only local almost isotropicity. To what extent we can
consider it to be homogeneous? It was the problem arised in 1960-s
by Yu.A.Volkov.

The main difficulty in solving the problem is that the usual proof
of Schur's theorem is based on Bianchi identities applied to
(\ref{e8}) where after summing we obtain $K_{,i} = 0$ hence $K =
{\rm const}$. However the fact that curvatures along different
directions are almost equal does not provide $K_{;i}$ to be small
enough. So, if we consider a space with almost equal curvatures
along all the directions as locally isotropic one, we cannot obtain
any isotropy. The goal of this section is to answer this
question assuming the local almost isotropicity to be the closeness
of results of all the measurements along any direction.

\begin{definition}\label{d4a} A spacetime region $M$ is called
$\epsilon$-small if for any $a,b \in M$

\[
\begin{array}{c}
|x^i(a) - x^i(b)| \le \epsilon  \cr
\rho(a,b) \le \epsilon
\end{array}
\]

\noindent where $\rho$ is a metric on $M$.
\end{definition}

\begin{definition}\label{d5} By an $(x^i)$-distance between points
$a,b\in M$ we shall call $\max_i\{|x^i(a) - x^i(b)|, \rho(a,b)\}$.
       The $\epsilon$-neighborhood of a point $a$ is the set of
all points $m$ of $M$ such that the $(x^i)$-distance between $m$
and $a$ does not exceed $\epsilon$. All geometric and kinematic
measurements, as it was already mentioned in section \ref{s2} are
reduced to the measurement of the metric, the distances and proper
time interval along trajectories of particles.
\end{definition}

\begin{definition}\label{d6} A spacetime region $M$ is
$\epsilon$-locally isotropic if for any $a\in M$ one can
define the action of an apropriate rotation group so that $a$ is
invariant under this action and all results obtained in
the $\epsilon$-neighborhood of $a$ coincide up to
$\lambda(\epsilon)$ with the results obtained after the action of
any element of the group.

\end{definition}

\begin{definition}\label{d7} In an analogous way we shall call a
region $\delta$-uniform if one can define an action of a transition
group on this region so that the results of any measurement on a
system of $N$ particles coincide with the precision
$\lambda(\epsilon)$ with analogous measurement on the system,
obtained from the first one after applying to it any element of the
group.

A region is called $\epsilon$-locally $\delta$-uniform if all the
above is valid for the measurement in  $\epsilon$-neighborhoods of
any two points of the region. Further we shall consider only
geodesically connected domains.

\end{definition}

The problem now is to find the least $\delta$ (over $\epsilon$,
$\delta$ and $L$) such that any $\epsilon$-locally
$\lambda$-isotropic region of the size $L$ is $\delta$-uniform.

\begin{theorem}\label{th2} Any $\epsilon$-locally
$\lambda$-isotropic region of the size $L$ is:

\begin{enumerate}
\item $C\epsilon$-locally  $\frac{CL\lambda}{\epsilon}$-uniform
\item $C(L/\epsilon)^2\lambda$-uniform
\end{enumerate}

\noindent where $C = {\rm const}$, and there are spaces for which
this evaluations cannot be diminished.
\end{theorem}

\paragraph{Sketch of the Proof.} 1). Since the local metric of
Riemannian space is close to Euclidean there exists some $c$ of
order 1 such that into the intersection of two $\epsilon$-neighborhoods
of two points on the distance $\epsilon$ one can inscribe a
$C\epsilon$-neighborhood.

2). Consider two $C\epsilon$-neighborhoods of arbitrary points
$x,y\in M$. Let $x^1,y^1$ be points on their frontiers (in
pseudo-Riemannian case we choose these points so that $xx^1$ and
$yy^1$ would be of the same kind). Since $M$ is geodesically
connected, $x^1$ and $y^1$ can be connected with a geodesic whose
$(x^i)$-length does not exceed $L$ (see Definition \ref{d5}). It
can be divided into $L/\epsilon$ parts of $(x^i)$-length of order
$\epsilon$. Then we approximate the geodesic by a broken line so
that the intervals $x^1z^1, z^1z^2, \ldots , z^ny^1$ have the
$(x^i)$-length $\epsilon$.

Now let us rotate $C\epsilon$-neighborhood of $x$ in order to
make it appears all inside the intersection of
$\epsilon$-neighborhoods of $x^1$ and $z^1$.

Then we rotate it around $z^1$, so that it gets to the
$\epsilon$-neighborhood of $z_2$ and so on until the
$\epsilon$-neighborhood of the point $y$. The results of all
measurements do not differ more than $\delta y\lambda$, hence the
results in neighborhoods of $x^1$ and $y^1$ do not differ by more
than $\lambda L/\epsilon$.

Here is an example when the evaluation cannot be refined: when
all differences of measurements are of the same sign i.e.
curvature monotonically varies from $x^1$ to $y^1$.

3). In a similar way we obtain the result for the global uniformity.
An interval of curve of length $\sim L$ is composed of $\sim
CL/\epsilon$ intervals of length $\sim C\epsilon$. The results of
measurements along this intervals are indistinguishable up to
$L\lambda/\epsilon$, hence the error in time or length measurement
along all the curve does not exceed $(CL/\epsilon)\cdot
(L\lambda/\epsilon) = C(L/\epsilon)^2\lambda$. This completes the
proof of Theorem \ref{th2}.

\section{Physical interpretation of the results and possible
applications}

If we perform in an $\epsilon$-small (see Definition \ref{d4a})
region a measurement with error $\lambda$ we know the metric
$g_{ij}\sim \delta\tau/\delta x^j$ up to $\lambda/\epsilon$,
the Cristoffel symbols $\Gamma^i_{jk}\sim \partial
g_{ik}/\partial x^j \sim \delta^2\tau/\delta x^{j2}$ up to
$\lambda/\epsilon^2$, and the curvature up to $\lambda/\epsilon^3$.
In any point we can choose a coordinate frame making
$\Gamma^i_{jk}$ zero, therefore measurements with error
$\lambda\sim \epsilon^2$ do not allow us to distinguish
non-isotropic case from a locally isotropic (and even from flat
one, i.e. any Riemannian space is $\epsilon$-locally
$C\epsilon^2$-isotropic). For the sake of such distinction the
error of the curvature must not exceed its value $K$, so the
relation $\lambda/K\epsilon^2$ characterize the relative error of
measurement of local {\em isotropy}.

The relative error of measuring the local isotropy is the
same.  Therefore local isotropy implies local uniformly with
$\epsilon/\lambda$ times greater error. Hence if $\lambda\sim
\epsilon^3$ we obtain (for small enough $\epsilon$) the
$\epsilon^2$-uniformity obtained above for any Riemannian metric.
To obtain non-trivial information on local uniformity one must have
$\lambda\le \epsilon^4$ which corresponds to the possibility of
exact enough measurement of curvature tensor and its derivatives,
viz. tidal forces and their spacetime gradients. Mathematically
it means that if both and $R_{ijkl}$ and $R_{ijkl;m}$ are almost
isotropic then the space is almost uniform.

Thus the physical result is the following: if accelerations
and tidal forces an locally isotropic then nothing can be said on
uniformity of the region.  However if gradients of tidal forces are
also isotropic then the space is locally uniform.

Imagine we verify the isotropy in a few points (e.g. close to
the Earth) - in general the same points as other points of space.
If it happens that in these points the space is $\epsilon$-locally
$\lambda$-isotropic then it is naturally to assume $\epsilon$-local
$\lambda$-isotropy everywhere and the results obtained give us the
evaluation of its homogeneity.


\begin{thebibliography}{99}

\bibitem{einstein}
Einstein, A.,
{\it The meaning of relativity,\/}
Princeton University Press,
Princeton, New Jersey, 1955 (4-th edition)

\bibitem{fock}
Fock, V.A.,
{\it Theorie von Raum, Zeit und Gravitation,\/}
Akademie-Verlag,
Berlin, 1960

\bibitem{c8}
Grib, A.A., and E.A.Poberii,
On the difference between conformal and minimal coupling in
general relativity,
{\it Helvetica Physica Acta,\/}
{\bf 68},
380--395,
1995

\bibitem{c7}
Lightman, A.P., W.H.Press, R.H.Price and S.A.Teukolski,
{\it Problem book in relativity and gravitation,\/}
Princeton University Press,
Princeton, New Jersey, 1975

\bibitem{c4}
Treder, H.-J.,
{\it Gravitationstheorie und \"Aquivalenzprinzip,\/}
Akademie-Verlag,
Berlin, 1971

\end{thebibliography}
\end{document}